# SOAP Serialization Performance Enhancement

## DESIGN AND IMPLEMENTATION OF A MIDDLEWARE

Behrouz Minaei

Computer Department
Iran University of Science and Technology
Tehran, Iran

Parinaz Saadat

Computer Department
Iran University of Science and Technology
Tehran, Iran

*Abstract—The most straightforward way to improve performance of any system is to define the bottlenecks and think of ways to remove them. Web services are the inseparable part of any web application, as a result enhancing performance of web services will have a great effect on the overall performance of the system. The most widely used communication protocol in the web services model, SOAP, is a simple protocol for the exchange of messages. The serialization of large SOAP responses is a major performance bottleneck in a SOAP message exchange.*

*Clearly, some web servers can expect to receive many similar messages for a particular web service as they share the same signature. The idea behind this paper is to avoid the redundant serialization stage of SOAP responses for request which have the same call parameters. The technique exploits the similarities between call parameters to improve web service Response Time by avoiding redundant serialization of the same response with the help of a middleware running on top of web server. The middleware will maintain a trie of incoming parameters for every set of current requests. This way request processing and serialization of the response of same requests will be done only once.*

*In a nutshell, to serialize only the different responses is the simplest way to avoid extra work done by a serializer. It might worth noting that although our approach is to utilize the exact repeating portion parameters, the middleware can be configured to apply changes made to the result set of response to the serialized response being maintained in a trie to generate valid results.*

***Keywords:Web Serivces,Performance,Middleware,Serialization***

## I. INTRODUCTION

Web service is a widely-used technology for exchanging data between applications and its scope of usage has widened even more in recent years.

In this paper, we describe the design and implementation of a server side middleware, which we call SEM (Serialization Enhancement Middleware). The idea is similar to the approach introduced in [1] but the implementation differs completely. In a nutshell, to serialize only the different responses is the simplest way to avoid extra work done by a serializer.

To illustrate the opportunity for performance enhancement, consider a web service for a search process in a search engine, each message arrives at the web services container to invoke a simple search. Aside from the search string, the only other difference in SOAP contents is the value of the "Content Length" field in the header. If this search engine has 500 request per second, messages arrived in succession, the SOAP server side would ordinarily have to parse and deserialize, process and then serialize the response all of them completely and independently of one another. Ideally, the server would be able to save time by recognizing that a large percentage of messages are completely the same, the result of these service calls needs to be serialized only once. Thus, the effectiveness of the Middleware optimization depends on the following factors:

- The percentage of same messages in an incoming message stream
- The percentage of similarity between different messages in an incoming message stream
- The overhead of message analysis.

## II. RELATED WORK

In [4, 3, 2] this problem is addressed on the sender's side, by avoiding serializing entire messages.The sender side of our SOAP implementation, called bSOAP, saves copies of outgoing message buffers, and tracks when the client code makes changes to the data sent in the messages. Only the changes are reconverted and rewritten into the outgoing message buffer template. The rest of the template remains unchanged from the previous send, avoiding serialization for that portion of the message. Our performance study indicates that this technique, called differential serialization (DS).

The approach in [1] describes the design and implementation of differential serialization's analogue on the server side, called differential deserialization (DDS). The idea is to avoid fully deserializing each message in an incoming stream of similar messages. Differential deserialization gets its name because the server-side parser deserializes the differences between an incoming message and a previous one.

SEM and DDS are completely separated and independent ideas and implementations; neither depends on the other for any portion of the performance enhancements; the two techniques represent very different realizations of the same high level idea; DS for sending SOAP data, and DDS for receiving it.SEM is a combination of DS and DDS, that is, it shares the idea behind both techniques. On the other hand serialization process will be improved but with a completely different implementation.

In general, SEM is more promising optimization technique than DS, because it is more applicable. DS only works if the same client sends a stream of similar messages. DDS can avoid deserialization of similar messages sent by multiple different clients while SEM does both.

### III. OUR PROPOSED SOLUTION

The first component responsible for handling requests in a Client-Server Model is the Web Server; therefore it would be the best candidate for hosting a middleware. Our approach is to implement a middleware to run on top of any web server (IIS,Apache,..) and act as the primary component for processing request.

By definition, Web services can be communicated with over a network using industry standard protocols, including SOAP. That is, a client and a Web service communicate using SOAP messages, which encapsulate the in and out parameters as XML. Fortunately, for Web service clients, the proxy class handles the work of mapping parameters to XML elements and then sending the SOAP message over the network [5].

This means that the SOAP message can be reached before and after Serialization/Deserialization process.

As the calling of web service methods has a unique signature, the probability of receiving requests with completely the same parameters for a service is so high. The idea behind this paper is to avoid the redundant serialization stage of SOAP responses for request which have completely the same parameters.

The approach will be even more efficient if a constraint is put on the method signature. Our researches show that the best case is the situation in which the method parameters are all string and the response is a result set.

### IV. DESIGN AND IMPLEMENTATION

This section describes SEM's design and implementation. Section 4.1 begins with a description of the middleware and discusses the algorithms used for comparison between SOAP messages. Section 4.2 then describes an optimization on the given approach. This is accomplished by a feature of the algorithm used for maintaining Soap Request Parameters which compares messages and considers overhead. Section 4.3 gives the alternative ways to enhance performance.

### A. The Middleware

Request comparison, analysis and processing consist of five main steps, each running in a different thread for maximum performance enhancement. Each step is described in detail as follows.

#### 1) Gathering Current Requests

In order to maintain web service statelessness, we concentrated on Current Requests on the web server. So we had to define the term "Current" in this context. In the implementation the term current requests is used for incoming messages during a predefined period of time. For the purpose of our implementation this predefined period of time was set to 2 milliseconds, a timer is activated and all incoming messages are collected in a dataset which is then passed to the next step each 2 milliseconds for further analysis

#### 2) Retrieving Parameter sequences

As soon as the Current Collection is ready it is passed to another thread, where the parameters are retrieved from each SOAP message and a sequence containing parameters is maintained for each Soap message.That way a large portion of messages can bypass the serialization phase if the message is totally the same.

In this phase parameters are retrieved from each Soap request in the Current Collection and a parameter sequence is generated for each request. If the sequence of parameters is duplicated, there is no need to do all the job of request processing and serialization of response for every single request. So all but one of duplicated sequences is ignored, but the id for each request is saved so that the serialized response can be sent for these duplicated requests.

| Sequence1 | Parameters of the first request in the Current Collection |
| Sequence2 | Parameters of the second request in the Current Collection |
| .. | .. |
| SequenceN | Parameters of the N-th request in the Current Collection |

Figure1.A List of Sequences of input parameters

#### 3) Comparing Parameter sequences

One of the most challenging issues of the approach is the algorithm by which the identical parameter sequences are detected. The simplest algorithm is to simply compare each of n sequence with other n-1 sequences in the collection. In many applications, it is necessary to determine the string similarity. Edit distance approach is a classic method to determine Field Similarity [8,9]. A well known dynamic programming algorithm is used to calculate edit distance with the time complexity O(nm). The Hamming distance also can be used.

A faster algorithm had to be chosen otherwise the comparison phase would be a bottleneck itself. So a data structure which satisfied the need was chosen. The opportunity for performance enhancement largely depends on the decision whether to use SEM or to continue with regular serialization. At this step based on the amount of similarity a decision has to be made whether to use the technique or take the regular steps of serialization combined with compression or JSON to enhance the performance.

In a more general case when some of the parameters are the same and some not. SEM utilizes an algorithm to track the amount of a parameter sequence occurrence and performs even better.

A **trie**, or prefix tree, is an ordered tree data structure that is used to store an associative array where the keys are usually strings. Unlike a binary search tree, no node in the tree stores the key associated with that node; instead, its position in the tree shows what key it is associated with. All the descendants of a node have a common prefix of the string associated with that node, and the root is associated with the empty string. Values are normally not associated with every node, only with leaves and some inner nodes that correspond to keys of interest[7].

So we chose to maintain a trie for the sequence collection that is to insert every parameter sequence in the trie. A sample code for doing so is as follows:

```
char[] charArray =
s.ToLower().ToCharArray();
TrieNode node = root;
foreach (char c in charArray)
{if (node.Contains(c))
node= node.GetChild(c);
else
{int n = Convert.ToByte(c) - TrieNode.ASCIIA;
TrieNode t = new TrieNode();
node.nodes[n] = t;
node= t;}
node.isEnd = true;}
```

Figure 2. Sample code for adding parameter sequences to the trie

By now, a collection of sequences of parameters is prepared. The next phase is to compare these sequences and find identical sequences so that the serialization step for identical sequences can be done just once. So trie lookup (and membership) can be used easily.

A key factor for choosing trie for detecting duplicate sequence of length $m$ takes worst case O($m$) time, in other words trie structure guarantees that no duplicate parameter sequence is maintained.

Besides tries require less space when they contain a large number of short strings, because the keys are not stored explicitly and nodes are shared between keys with common initial subsequences.

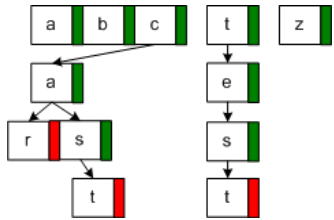

Figure3. A Trie.Nodes are represented by an array of pointers

So at the end of each sequence just before tagging the end of the sequence, it is clear if the request is duplicated, if so that request is marked as the duplicate so that when the response of that particular request is ready we can send the response for that too.

```
char[] charArray = s.ToLower().ToCharArray();
TrieNode node = root;
bool contains = true;
foreach (char c in charArray)
{node = Contains(c, node);
if (node == null)
{contains = false;
break;}}
if((node == null) ||(!node.isEnd))
contains = false;
```

Figure 4. Sample code for searching if a parameter sequences is already added to the trie

The following are two main advantages of tries:

- Looking up keys is faster. Looking up a key of length $m$ takes worst case O($m$) time. A BST performs O($\log(n)$) comparisons of keys, where $n$ is the number of elements in the tree, because lookups depend on the depth of the tree, which is logarithmic in the number of keys if the tree is balanced. Hence in the worst case, a BST takes O($m \log n$) time. Moreover, in the worst case $\log(n)$ will approach $m$. Also, the simple operations tries use during lookup, such as array indexing using a character, are fast on real machines[7].

- A trie can provide an alphabetical ordering of the entries by key.

Tries do have some drawbacks as well. Tries can be slower in some cases than hash tables for looking up data, especially if the data is directly accessed on a hard disk drive or some other secondary storage device where the random access time is high compared to main memory.

*4) Regular Processing*

After this phase the requests are processed, that is every single request, plus one out of n identical request, are sent to the web server, where they are deserialized and processed.

*5) Sending Duplicate Responses*

When the response of that request is ready it is sent to all other identical requests as well.

*B. Yet another Optimization*

In order to gain the maximum performance enhancement, the serialized response having the most number of requests in each Current Collection is also maintained in another trie,so that each time another request with those parameters arrives, the response can be generated by some inserts and updates.

There is no need to be worry about memory limits as when the trie is mostly static and when the trie nodes are not keyed by node specific data (or if the node's data is common) it is possible to compress the trie representation by merging the common branches.

The result of such compression may look similar to trying to transform the trie into a directed acyclic graph (DAG), because the reverse transform from a DAG to a trie is obvious and always possible.

### C. Alternatives

There is a tradeoff between overhead of comparing messages and the performance enhancement gained. As a result, the portion of the same parameter sequence determines whether it is worth taking advantage of our approach. That is, there exists an opportunity for switching to regular Serialization/Deserialization when requests are not the same.

Consequently, the process of detecting when to utilize the technique is twofold. Firstly, we measure the percentage of same requests per Current Collection. Secondly we must ensure that the overhead regarding comparison and analysis worth it.

Imagine the case when the parameter values differ totally. In this case the overhead of our approach would be so high, so an alternative approach is chosen. This alternative approach could be using other format for serialization/deserialization such as JSON[1], using techniques to compress the SOAP message, etc.

### V. EXPERIMENTAL SETUP AND ANALYSIS

We ran all performance tests on a single Pentium 4 3.00 GHz machine with 3.24 GB of RAM, and a 100GB SATA Drive.

In order to be able to simulate any situation a Load Test Generator was also developed. This application generates SOAP request and calls a web service method in the following situations:

- Simulation of X Concurrent Requests per second (with threads) from Y Clinets
- Simulation of X Serial Requests per second from 1 client
- A Monitoring tool for monitoring the web server's Performance Counters such as Byte Received/Sec, Byte Send/Sec, Total Bytes/Sec, and Connection Attempts/Sec.

Multiple situations were tested so that the results can be compared for a better conclusion. In each situation the amount of similarity between web service calls and the number of web service call were the factors indicating the performance enhancement or degradation.

The method was the same for all situations, a request to a web service was simulated, and every step was monitored till the response was ready.

- Situation A simulated 1000 concurrent requests for the same web service, each with completely different parameter values for the call.
- Situation B simulated 1000 concurrent requests for the same web service, with completely the same parameter values for the call.
- Situation C simulated 1000 concurrent requests for the same web service, with 50% the same parameter values for the call.

The monitoring tool then showed the results of each.

The middleware, running on top of IIS 7.0, read a sequence of "incoming" SOAP messages from the load test generator and passes them to the comparison algorithm module where call parameters of SOAP message were retrieved. At this step based on the amount of similarity a decision was made, whether to use the technique or take the regular steps of serialization combined with compression or JSON to enhance the performance.

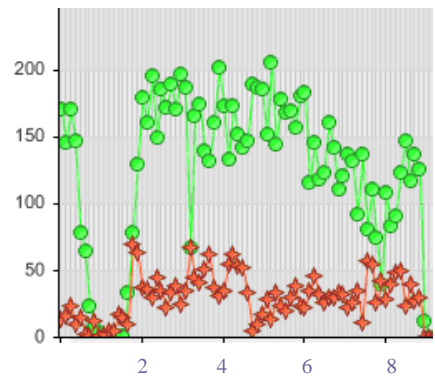

Figure5. The performance enhancement of middleware when the percentage of similar calls is 70%

As shown in figure 3, the number of requests per second is nearly three times more, when the percentage of the same calls is 70.

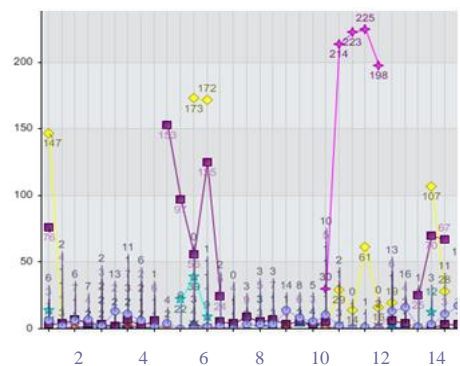

Figure5. Parameters fetched from the parameter collection

---

[1] Java Script Object Notation

As mentioned before, a collection of parameters is maintained and call simulator uses this collection to simulate server calls. When there are more similar parameter sequences, the effectiveness of SEM on performance is significantly higher.

Table 1 shows the final results of the implementation. In this table, response time of requests is shown in milliseconds. Clearly, sequences of identical messages do not represent a realistic scenario. The values reported in this section show that as the percentage of parameter sequence similarity increases response time of request processing using SEM gets better.

TABLE I.    THE DEGREE OF PARAMETER SEQUENCE RESEMBLANCE LEADS TO DIFFERENT RESPONSE TIMES

| Parameter Sequence resemblance | Trie Depth | | | | |
|---|---|---|---|---|---|
| | 15 | 25 | 35 | 65 | 75 |
| 10% | 2.89 | 28.7 | 71.5 | 144 | 288 |
| 20% | 0.83 | 8.24 | 17.7 | 42.7 | 85 |
| 30% | 0.70 | 6.93 | 16.0 | 35.4 | 71.3 |
| 40% | 0.67 | 5.67 | 13.6 | 28.6 | 63.9 |
| 50% | 0.57 | 1.72 | 4.29 | 8.03 | 55.4 |
| 60% | 0.56 | 1.32 | 3.23 | 5.89 | 16.7 |
| 70% | 018 | 0.75 | 2.11 | 3.69 | 12.5 |
| 80% | 0.13 | 0.67 | 1.85 | 2.95 | 7.72 |
| 90% | 0.05 | 0.51 | 0.78 | 1.54 | 3.55 |
| 100% | 0.04 | 0.21 | 0.60 | 1.27 | 2.91 |

Another point is that the depth of trie, parameter sequence length, has also a dramatic effect on performance enhancement.

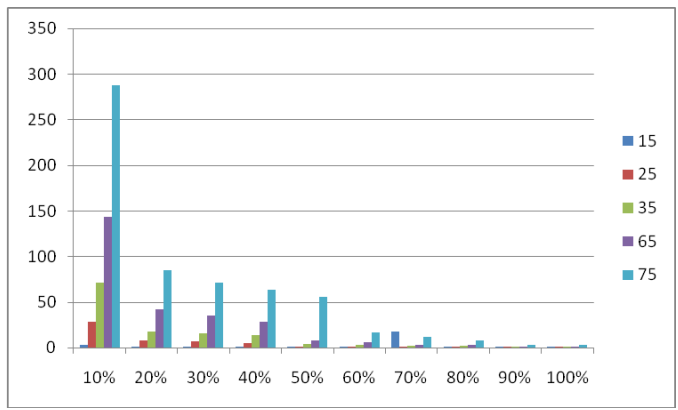

Figure6. The effect of trie depth on response time

Figure 6 illustrates the effect of tire depth on response time for requests with different percentage of similarities. As shown in the figure deeper tries cause less performance but this impact can be lessened as the percentage of similarity between concurrent messages increases. It might worth noting that by trie depth is related to the length of call parameters.

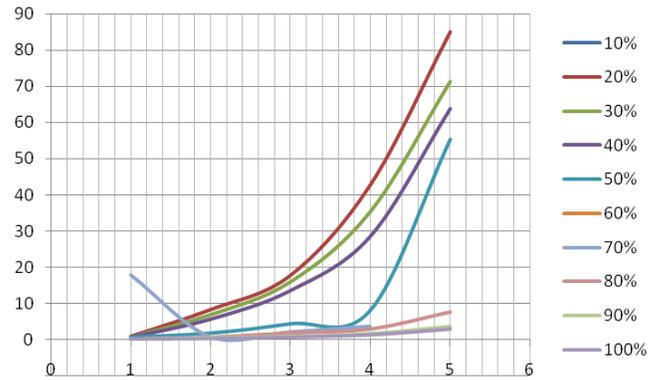

Figure7. The effect of parameter similarity on response time

As shown in figure 7 parameter similarity has a great impact on service call response time. As shown in the figure as the similarity of call parameters between concurrent messages increases better response time is gained. The optimal condition is when there are exact same call parameters between concurrent messages.

## VI.    SUMMARY AND FUTURE WORK

Serialization Enhancement Middleware (SEM) is a Middleware running on top of web server to take advantage of similar Soap requests on a web server for a particular web service. This way a large portion of responses can bypass the serialization phase if the message is totally the same.

Current requests for incoming messages during a predefined period of time are collected in a dataset which is then passed to the next step Each 2 seconds for further analysis. Next parameters are retrieved from each Soap request and a parameter sequence is generated for each request. If the sequence of parameters is duplicated, there is no need to do all the job of request processing and serialization of response for every single request.

Then a trie is maintained for the sequence collection that is to insert every parameter sequence in the trie. A key factor for choosing trie for detecting duplicate sequence of length *m* takes worst case $\underline{O}(m)$ time,in other words trie structure guarantees that no duplicate parameter sequence is maintained.

After this phase the distinct requests, plus one out of n identical request are deserialized and processed. When the response of that request is ready it is sent to all other identical requests as well.

Although our approach is to utilize the exact repeating portion parameters, one optimization is to enable the Middleware so that it can be configured to apply changes made to the result set of response to the serialized responses being maintained in a trie to generate valid results. But this can also lead to a larger percentage of time spent for the comparison and analysis phase